# On the effective conductivity of composite materials

Yuri Kornyushin

*Maître Jean Brunschvig Research Unit, Chalet Shalva, Randogne, CH-3975, Switzerland*

A composite conductive material, which consists of fibers of a high conductivity in a matrix of low conductivity, is discussed. The effective conductivity of the system considered is calculated in Clausius-Mossotti approximation. Obtained relationships can be used to calculate the conductivity of a matrix, using experimentally measured parameters. Electric fields in the matrix and the inclusions are calculated. It is shown that the field in a low-conductivity matrix can be much higher than the external applied one.

In a homogeneous conductor the density of the electric current, $J$, is a product of specific electric conductivity, $\sigma$, and electric field intensity, $E$ [1]. In inhomogeneous conductor averaged over the volume of a sample electric current, $\langle J \rangle$, is a product of effective specific electric conductivity, $\sigma_e$, and averaged over a sample volume electric field $\langle E \rangle$, produced by homogeneously applied external electric field (see, e. g. [2]). The problem of calculation of effective specific electric conductivity in inhomogeneous sample is one of the extreme mathematical difficulties [2]. One of the successful approximations in this field is Clausius-Mossotti approximation. In [2] Clausius-Mossotti approximation is applied, in particular, to the modeling of effective electric conductivity of a sample, containing arbitrary amount of ellipsoidal conductive inclusions of the same shape and orientation in a conductive matrix. It is assumed also that the current and the electric field in each inclusion and in the matrix are homogeneous ones, but they are different in different inclusions and in the matrix. It is obtained in [2] that

$$\sigma_e = \sigma_m \{1 - (1-n)\Sigma f_k(\sigma_m - \sigma_k)/[(1-n)\sigma_m + n\sigma_k]\}/\{1 + n\Sigma f_k(\sigma_m - \sigma_k)/[(1-n)\sigma_m + n\sigma_k]\}, \quad (1)$$

where $\sigma_m$ is the (effective) specific electric conductivity of a matrix, and $n$ is a form factor of the inclusion in one of the main directions. It coincides with so called depolarization factor [1, 2]. The inclusions were assumed in [2] to be of different types, made of materials with different conductivities $\sigma_k$. In Eq. (1) $f_k$ is the volume of $k$-type inclusions, divided by the volume of a sample (that is $f_k$ is the volume fraction of $k$-type of inclusions).

Now let us consider a sample with inclusions of one type. Then Eq. (1) yields:

$$\sigma_e = \sigma_m[f\sigma + (1-n)(1-f)\sigma_m + n(1-f)\sigma]/[(1-n)\sigma_m + n\sigma - nf(\sigma - \sigma_m)]. \quad (2)$$

For volume fraction of the inclusions $f$ close to 1 and when the conductivity of the matrix $\sigma_m$ is essentially smaller than that of the inclusions $\sigma$, $(1-n)(1-f)\sigma_m$ can be neglected in the numerator of Eq. (2) comparative to $f\sigma$. The term $n(1-f)\sigma$ also could be neglected in the numerator of Eq. (2). In the denominator of Eq. (2) $\sigma_m$ can be neglected comparative to $\sigma$. After that we have:

$$\sigma_e = f\sigma\sigma_m/[(1-n)\sigma_m + (1-f)n\sigma]. \quad (3)$$

When in the denominator of Eq. (3) $(1 - n)\sigma_m$ can be neglected comparative to $(1 - f)n\sigma$, Eq. (3) yields:

$$\sigma_e = f\sigma_m/(1 - f)n. \qquad (4)$$

It follows from Eq. (3) also that

$$\sigma_m = (1 - f)n\sigma\sigma_e/[f\sigma - (1 - n)\sigma_e]. \qquad (5)$$

This equation can be used to calculate $\sigma_m$ from measured values of $\sigma, f, n$, and $\sigma_e$.

Let us regard a case of inclusions being conductive fibres of a shape of long cylinders of the length $l$ and diameter $d$. We assume that the volume fraction $f$ is close to 1. Depolarisation factor of a long cylinder in the direction of the long axis can be approximated as one of a very long prolate spheroid with axes $l$, and $d$. For a very long prolate spheroid [3]

$$n = (d/l)^2[\ln(2l/d) - 1]. \qquad (6)$$

This quantity is an extremely small one. In the direction, perpendicular to the axis of a fibre, the depolarisation factor is very close to 0.5.

From the symmetry of the regarded objects follows that when the direction of the applied field is changed by $180^{\circ}$, the direction of the measured current is also reverted. But the ratio of the measured field and current is not changed. From this follows that measured effective conductivity can be only an even function of applied electric field.

Let us calculate fields in the matrix and in the inclusions. Let $E_m$ denotes average field in the matrix and $E_i$ denotes that one in the inclusions. As the average field in the sample is $\langle E \rangle$, we have

$$\langle E \rangle = fE_i + (1 - f)E_m. \qquad (7)$$

In the case of superconductive inclusions $E_i = 0$ and Eq. (7) yields

$$E_m = \langle E \rangle/(1 - f). \qquad (8)$$

This field acts on the inclusions. When $f$ is close to unity, $E_m$ is essentially larger than $\langle E \rangle$.

To calculate the fields mentioned in Clausius-Mossotti approximation we need also following relationship:

$$\langle J \rangle = \sigma_e\langle E \rangle = f\sigma E_i + (1 - f)\sigma_m E_m. \qquad (9)$$

It follows from Eqs. (7) and (9) that at any volume fraction $f$ of arbitrary value

$$E_m = (\sigma\langle E \rangle - \langle J \rangle)/[(1 - f)(\sigma - \sigma_m)], \qquad (10)$$

$$E_i = (\langle J \rangle - \sigma_m\langle E \rangle)/[f(\sigma - \sigma_m)]. \qquad (11)$$

From Eqs. (5), (10) and (11) follows that

$$E_m = \{(\sigma - \sigma_e)[f\sigma - (1 - n)\sigma_e]/\sigma(1 - f)(f\sigma - \sigma_e + fn\sigma_e)\}\langle E \rangle, \qquad (12)$$

$$E_i = \{\sigma_e[f\sigma - (1-f)n\sigma - (1-n)\sigma_e]/f\sigma(f\sigma - \sigma_e + fn\sigma_e)\}\langle E\rangle. \tag{13}$$

Eqs. (12, 13) are applicable when volume fraction $f$ is close to unity.

As $E_m$ is the field, acting on the inclusions, the dependence $\sigma_m$ on electric field should be presented as $\sigma_m = \sigma_m(E_m)$.

I would like to thank Eugene B. Gordon of the Institute of Problems of Chemical Physics, RAS, Chernogolovka for drawing my attention to the interesting small problem, discussed in this paper.

# Об эффективной проводимости композитных материалов

## Ю. В. Корнюшин

*Maître Jean Brunschvig Research Unit, Chalet Shalva, Randogne, CH-3975, Switzerland*

Обсуждаются композитные материалы, состоящие из хорошо проводящих волокон и плохо проводящей матрицы. Эффективная удельная проводимость рассматриваемой системы вычислена в приближении Клаузиуса-Моссотти. Полученные соотношения могут быть применены для вычисления удельной проводимости матрицы, используя экспериментально измеренные другие величины. Рассчитаны электрические поля в матрице и включениях. Показано, что поле в плохо проводящей матрице может значительно превосходить внешнее приложенное поле.

В однородном проводнике плотность электрического тока, $J$, равна произведению удельной электропроводности, $\sigma$, и напряженности электрического поля, $E$ [1]. В неоднородном проводнике усреднённая по объёму образца плотность электрического тока, $\langle J\rangle$, равна произведению удельной эффективной электропроводности, $\sigma_e$, и усреднённой по объёму образца напряженности электрического поля, $\langle E\rangle$, возникающего в результате раномерного приложения внешнего электрического поля (смотри, например, [2]). Задача о вычислении эффективной удельной электропроводности неоднородного образца является проблемой черезвычайной математической сложности [2]. Приближение Клаузиуса-Моссотти успешно применяется в этой области. В [2] приближение Клаузиуса-Моссотти применено, в частности, для моделирования эффективной удельной электропроводности образца, содержащего произвольное число эллипсоидальных проводящих включений одной и той же формы и ориентации в проводящей матрице. Предполагается также, что плотности тока и электрические поля в каждом включении и в матрице являются однородными, но они различны в различных типах включений и в матрице (кусочно-однородное приближение). В [2] получено, что

$$\sigma_e = \sigma_m\{1 - (1-n)\Sigma f_k(\sigma_m - \sigma_k)/[(1-n)\sigma_m +$$

$$n\sigma_k]\}/\{1 + n\Sigma f_k(\sigma_m - \sigma_k)/[(1-n)\sigma_m + n\sigma_k]\}, \quad (1)$$

где $\sigma_m$ – (эффективная) удельная электропроводность матрицы, $n$ – фактор формы включения вдоль одной из главных осей. Этот фактор совпадает с фактором деполяризации [1, 2]. В [2] рассматривались включения различных типов, имеющих различные электропроводности $\sigma_k$. В (1) $f_k$ – это отношение объёма включений типа $k$ к объёму образца (то есть $f_k$ – это объёмная доля включений типа $k$).

Рассмотрим образец, содержащий включения одного типа. В этом сучае из (1) следует, что

$$\sigma_e = \sigma_m[f\sigma + (1-n)(1-f)\sigma_m + n(1-f)\sigma]/[(1-n)\sigma_m + n\sigma - nf(\sigma - \sigma_m)]. \quad (2)$$

В случае, когда объёмная доля включений $f$ близка к 1 и проводимость матрицы $\sigma_m$ существенно меньше проводимости включений $\sigma$, слагаемым $(1-n)(1-f)\sigma_m$ в числителе (2) можно пренебречь по сравнению со слагаемым $f\sigma$. В числителе (2) также можно пренебречь слагаемым $n(1-f)\sigma$. В знаменателе (2) можно пренебречь $\sigma_m$ по сравнению с $\sigma$. В результате получаем:

$$\sigma_e = f\sigma\sigma_m/[(1-n)\sigma_m + (1-f)n\sigma]. \quad (3)$$

Когда в знаменателе (3) можно пренебречь $(1-n)\sigma_m$ по сравнению с $(1-f)n\sigma$, из (3) следует, что

$$\sigma_e = f\sigma_m/(1-f)n. \quad (4)$$

Из (3) следует также, что

$$\sigma_m = (1-f)n\sigma\sigma_e/[f\sigma - (1-n)\sigma_e]. \quad (5)$$

Это уравнение можно использовать для вычисления $\sigma_m$ из измеряемых величин $\sigma$, $f$, $n$ и $\sigma_e$.

Теперь рассмотрим случай, когда включения являются проводящими волокнами в форме длинных цилиндров длины $l$ и диаметра $d$. Объёмная доля включений в образце $f$ предполагается близкой к 1. Фактор деполяризации длинного цилиндра вдоль длинной оси можно считать приблизительно равным таковому сильно вытянутого сфероида с осями $l$ и $d$. Для очень длинного вытянутого сфероида [3]

$$n = (d/l)^2[\ln(2l/d) - 1]. \quad (6)$$

Эта величина очень мала. В перпендикулярном к оси цилиндра направлении фактор деполяризации весьма близок к 0,5.

Из симметрии рассматриваемых объектов следует, что если изменить направление внешнего приложенного поля на $180^o$, направление тока также изменится на противоположное. Однако отношение измеряемого тока к внешнему полю останется неизменным. Из сказанного следует, что измеряемая эффективная электропроводность рассматриваемых объектов может быть только чётной функцией внешнего приложенного поля.

Сейчас вычислим поля в матрице и включениях. Пусть $E_m$ обозначает среднее поле в матрице и $E_i$ обозначает среднее поле во включениях. Так как среднее поле в образце есть $\langle E \rangle$, мы имеем

$$\langle E \rangle = fE_i + (1-f)E_m. \tag{7}$$

В случае сверхпроводящих включений $E_i = 0$ и из (7) следует, что

$$E_m = \langle E \rangle/(1-f). \tag{8}$$

Именно это поле действует на включения. В случае близких к единице $f$ величина $E_m$ значительно превосходит $\langle E \rangle$.

Для того, что-бы рассчитать упомянутые поля в приближении Клаузиуса-Моссотти, необходимо ещё одно соотношение:

$$\langle J \rangle = \sigma_e \langle E \rangle = f\sigma E_i + (1-f)\sigma_m E_m. \tag{9}$$

Из уравнений (7) и (9) следует, что при произвольной величине фактора заполнения $f$:

$$E_m = (\sigma \langle E \rangle - \langle J \rangle)/[(1-f)(\sigma - \sigma_m)], \tag{10}$$

$$E_i = (\langle J \rangle - \sigma_m \langle E \rangle)/[f(\sigma - \sigma_m)]. \tag{11}$$

Из (5), (9) и (10) следует, что

$$E_m = \{(\sigma - \sigma_e)[f\sigma - (1-n)\sigma_e]/\sigma(1-f)(f\sigma - \sigma_e + fn\sigma_e)\}\langle E \rangle, \tag{12}$$

$$E_i = \{\sigma_e[f\sigma - (1-f)n\sigma - (1-n)\sigma_e]/f\sigma(f\sigma - \sigma_e + fn\sigma_e)\}\langle E \rangle, \tag{13}$$

Уравнения (12, 13) применимы в случае, когда объёмная доля включений $f$ близка к единице.

Так как $E_m$ является полем, действующим на включения, зависимость $\sigma_m$ от электрического поля должна быть представлена как $\sigma_m = \sigma_m(E_m)$.